\documentstyle[aps,epsf,pre,twocolumn]{revtex}

\draft 

\begin{document}

\title{Direct observation of twist mode in electroconvection in I52}

\author{Michael Dennin}
\address{Department of Physics and Astronomy}
\address{University of California at Irvine}
\address{Irvine, CA 92697-4575.}

\date{\today}

\maketitle

\begin{abstract}

I report on the direct observation of a uniform twist mode
of the director field in
electroconvection in I52. Recent theoretical work suggests
that such a uniform twist mode of the director
field is responsible for a number of secondary
bifurcations in both electroconvection and
thermal convection in nematics. I show here evidence that
the proposed
mechanisms are consistent with being the source of
the previously reported SO2 state
of electroconvection in I52. The same mechanisms also contribute to
a tertiary Hopf bifurcation that I observe in
electroconvection in I52. There are
quantitative differences between the experiment
and calculations that only include the twist mode.
These differences suggest that a complete description
must include effects described by the weak-electrolyte
model of electroconvection. 
\end{abstract}


\section{Introduction}

When a spatially extended system is driven far from
equilibrium, a series of transitions occurs
as a function of the external driving force,
or control parameter. The initial transition is typically
from a spatially uniform state to a state with periodic
spatial variations, called a pattern \cite{REV}. One can
distinguish two general classes of pattern forming systems:
isotropic and anisotropic. In isotropic systems, because there is
no intrinsic direction in the system, the initial wavevector
of the pattern can have any orientation. For anisotropic
systems, the uniform state of the system has a special
axis and there are at most two degenerate initial
wavevectors for the pattern.
Electroconvection in nematic liquid crystals has become a
paradigm for the study of pattern formation in anisotropic
systems \cite{LCREV,ECART}.

Nematic liquid crystals are fluids
in which the molecules possess orientational order \cite{GCLC}.
The axis along which the molecules are aligned on average 
is referred to as the director. For electroconvection,
a nematic liquid crystal is placed between two glass
plates. The plates are treated so that there is a
uniform alignment of the director parallel to the
plates, i.e. planar alignment. The plates are also
coated with a transparent conductor, and the liquid
crystal is doped with an ionic impurity. An ac voltage
is applied perpendicular to the director. Above
a critical value $V_c$ for the voltage, a pattern develops that
consists of a periodic variation of the director and
charge density with a corresponding convective flow of the
fluid. Many of the interesting patterns in electroconvection
are the result of {\it oblique rolls}. Oblique rolls
refer to patterns where the wavevector forms a nonzero
angle $\theta$ with respect to the initial alignment of the director.
Because the director only defines an axis, for each value
of $\theta$ and wavenumber $q$, there are two degenerate
states corresponding to wavevectors at the angles $\theta$ and
$-\theta$. These states are referred to as zig and zag,
respectively.

Electroconvection has been extensively studied
experimentally \cite{ECART}. However,
despite a relatively early identification of the basic instability
mechanism \cite{TC67,H69}, a detailed, quantitative description of
the rich array of patterns has only recently emerged. The first step
in this development was the elucidation of the standard model
of electroconvection \cite{SMREFS,Zi91}. The linear stability
analysis and weakly nonlinear analysis presented in
Ref.~\cite{SMREFS} accurately describes
electroconvection at relatively high electrical
conductivities and thick samples. However, it fails to account
for
traveling patterns, i.e. a Hopf bifurcation, that are
observed in thin samples and at low
sample conductivity \cite{RRFJS88,RRS89}. Also, the original
weakly nonlinear analysis of the standard model does
not explain the experimentally observed
``abnormal'' rolls\cite{PDRKPBR97,RZKR98}.
Recently, these two phenomena have been explained by
independent theoretical extensions of the standard model
that are described below.

First, the weak-electrolyte model (WEM) \cite{TK94,DTKAC95} is an
extension of the standard model that
treats the charge density as a dynamically active field and
is able to explain the Hopf bifurcation. 
Second, within the framework of the standard model, secondary and
further bifurcations have been assessed with a fully
nonlinear Galerkin calculation \cite{PDRKPBR97,PP99}. In particular,
this work helped elucidate the decisive role of a homogeneous in-plane twist
of the director in the bifurcation to abnormal rolls \cite{PDRKPBR97,RZKR98}.
The general features of this fully nonlinear calculation can be
reproduced by an extended weakly nonlinear analysis. This
analysis extends
previous treatments of the standard model by including the
homogeneous twist as a dynamically active mode \cite{PP99,PR98a},
and I will refer to it as the ``twist-mode model''.
For certain cases, the twist-mode model even provides
a semiquantitative or quantitative description of the dynamics.
This analysis applies to both
electroconvection and thermal convection
in nematics \cite{PP99}.

Currently, a weakly nonlinear
theory that includes both the WEM
effects and the twist mode remains undeveloped.
The appropriate merging of the twist-mode model and the WEM is
essential for systems with traveling oblique rolls, where
both effects can be important. For example, even
though electroconvection in I52
provided the first quantitative
confirmation of the WEM at the linear level \cite{DTKAC95},
the patterns in I52 are dominated by oblique rolls \cite{MYTH}.
Therefore, the homogeneous in-plane twist is present and may
be important at the weakly nonlinear level. The main question
is: does this mode need to be included as an additional
active mode in an
extended weakly nonlinear analysis, or
is the WEM model sufficient?
In particular, this question is crucial for
two patterns that are observed
in I52 where the dynamics depends on the interaction
between traveling oblique rolls: spatiotemporal chaos at
onset \cite{DAC96a}
and localized states known as ``worms'' \cite{DAC96b}. Currently,
qualitative features of these patterns have been reproduced in the
context of a weakly nonlinear analysis based on the WEM \cite{TK98}.
However, if the twist-mode is an active mode, it must be included for
any quantitative comparison to work.

The results reported here are a first step in determining
the relative importance of the in-plane homogeneous twist of the director
in electroconvection in I52.
For these initial experiments, the contribution of WEM effects
were minimized by focusing on relatively high conductivities.
This work includes a more detailed and quantitative study of
the SO2 state that is first reported in Ref.~\cite{DCA98}. The SO2
state consists of the superposition of rolls with
wavevector {\bf q} with rolls with wavevector {\bf k}. 
The angle between the wavevectors ranged from 72$^{\circ}$ to
90$^{\circ}$, depending on the parameter values.
In this paper, I will refer to the initial oblique roll
wavevector as {\bf q} and any subsequent
wavevector that grows as a result of a secondary instability
as the dual wavevector {\bf k}. Also, wavevectors with a positive
angle relative
to the undistorted director will be referred to as zig-type,
and wavevectors with a negative angle with respect to
the undistorted director will be referred to as a
zag-type. In general, because of the two-fold degeneracy,
the initial wavevector {\bf q} can either be zig-type
or zag-type.
 
In Ref.~\cite{PP99}, it has been proposed that the SO2
state is an example of the
bimodal varicose state. I will show that this
association is correct. 
Also, I report on direct measurements of a twist
mode of the director field in I52 for the oblique roll
states and the SO2 states.
These measurements confirm that the director-wavevector frustration
mechanism proposed in Ref.~\cite{PP99} is the source of the
bimodal instability that results in the
SO2 state. It has been predicted in Ref.~\cite{PP99} that the bimodal
state can experience a Hopf bifurcation to a time periodic state.
In this state, the amplitudes of the {\bf q}, {\bf k}, and homogeneous
twist modes all oscillate in time. This state is referred to as
the {\it oscillating bimodal varicose} \cite{PR98b}.
This state has been observed indirectly in thermal
convection \cite{PR98b}, where only the oscillations of
the {\bf q} and {\bf k} modes were observed. 
I report on observations of this Hopf bifurcation
in electroconvection. In particular, I have directly measured
the oscillations of the homogeneous twist mode in addition to the
oscillations of the {\bf q} and {\bf k} modes.

The sequence of bifurcations reported here is in perfect agreement
with the twist-mode model. Also, the measured angle between the {\bf q}
and {\bf k} modes agree with predictions of the twist-mode model.
However, the location of the bifurcation to the bimodal varicose
and the oscillating bimodal varicose in not quantitatively
described by the twist-mode model. This is easily attributed to
two facts. First, the material parameters are not completely known.
Second, the WEM effects are still present at some level in the
experimental system and
have not yet been incorporated into
the model.

The rest of the paper
is organized as follow. Section II describes the experimental
details. Section III presents the experimental results, and
Sec. IV discusses the comparison between the results and the
twist-mode model.

\section {Experimental Details}

I used the nematic liquid crystal I52 \cite{FGWP89}
doped with 5\% by weight
molecular iodine. Commercial cells were obtained from EHC, Ltd
in Japan \cite{EHCO}.
The cells consisted of two pieces of glass coated
with transparent electrodes of indium-tin oxide (ITO). The surfaces
were treated with rubbed polymers to obtain uniform planar
alignment. The initial alignment direction will be referred to
as the x-axis, and the z-axis is taken perpendicular to the
glass plates. The cell spacing was 23 $\mu$m and the electrodes
were 1 cm x 1 cm. The cells were filled by capillary action and
sealed with five minute epoxy.

The cells were placed in a temperature control block.
The temperature was maintained constant to $\pm 2\ {\rm mK}$.
In order to study a range of parameters,
four different operating temperatures were used:
35 $^\circ$C, 42 $^\circ$C, 45 $^\circ$C, and 55 $^\circ$C.

The system used to image the patterns is shown in Fig.~1.
It is a standard shadowgraph setup that is modified to allow
for the direct observation of in-plane
twist modes of the director using a $\lambda/4$
plate and analyzer. Two main orientations of the 
polarizer, $\lambda/4$ plate, and analyzer were used.
In Fig.~1a, the polarizer and analyzer are shown parallel
to each other and the initial alignment of the director.
With this geometry, only the transmission of extraordinary
light is observed. Minus the $\lambda/4$ plate and analyzer,
this is the standard shadowgraph setup
for electroconvection. The light is focused
by the director variation in the x-z plane, and an image of the
pattern is obtained \cite{RHWR89}.
In the arrangement shown in Fig.~1b, the polarizer is
oriented perpendicular to the initial alignment of the director.
In this case, only ordinary incident light is present, and
the contrast due to the x-z periodic variation is avoided.

In both cases, the $\lambda/4$
plate at $45^\circ$ to the undistorted director orientation is used
to discriminate domains of positive and negative x-y director twist.
Figure 1c is a top view of the cell and illustrates the definition
of positive and negative rotation. The optical setup used here
is analyzed in detail in Ref.~\cite{ASR99} for a thicker cell and
different liquid crystal. However, the qualitative features hold true
in this case. For the orientations shown in Fig.~1b, regions of
negative twist have an enhanced transmission of light
relative to regions of positive twist.
Rotating the $\lambda/4$
plate to $-45^\circ$ yields equivalent information, but with
complementary intensities.
With the polarizer aligned parallel to the undistorted director,
the dominant feature of the image is the focusing due to
the periodic x-z variation of the director. However, because we also
used the $\lambda/4$ plate in this case, there is a modification of the
overall intensity when a homogeneous twist is present. In this orientation,
because the polarizer and analyzer are both aligned with the undistorted
director, a positive twist of the director
has an enhanced transmission of light relative to
a negative twist of the director when the $\lambda/4$ plate is at
$+45^\circ$. Therefore, in both setups, the sign of the twist amplitude
can be determined. In the future, a detailed calculation along the
lines presented in Ref.~\cite{ASR99} is required for quantitative
measurements of the twist amplitude.

\begin{figure}[htb]
\epsfxsize = 3.2in
\centerline{\epsffile{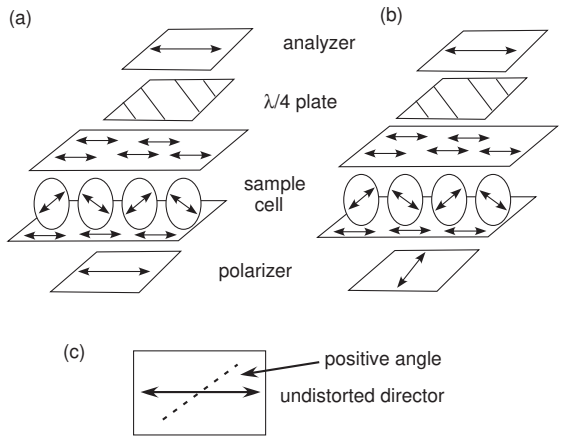}}
\caption{Schematic drawing of the optical setup.
(a) Setup used to image the periodic roll structure. In this case,
the polarizer is aligned with the undistorted director axis and
extraordinary light is used. (b) Setup used to image the in-plane
twist of the director. In this case, the polarizer is rotated
$90^\circ$ with respect to the undistorted director and
ordinary light is used. In addition to rotation of the polarizer,
both the $\lambda/4$ plate and the analyzer are free to rotate.
(c) A top view of the
cell that shows the undistorted director and the definition
of positive angles used for both the in-plane
director twist and the orientation of the $\lambda/4$ plate.
This top view is the same orientation as used for all of the
images presented here.}
\end{figure}

The images were taken using a standard CCD camera and digitized with
an 8-bit framegrabber. All of the images have both a background
subtraction and background division performed. The background
subtraction was done to remove the effective mean from the image that is
due to the fact that the camera digitized images on
a 0 to 255 scale. This was necessary to enable detection of the uniform
twist mode by Fourier techniques. The presence of a uniform twist shifts
the mean intensity of the image and shows up as changes in the amplitude
of the zero wavevector. Without the subtraction step, the zero wavevector
peak is dominated by the mean caused by the digitizing process.

The electrical conductivity of the sample ranged between
$1 \times 10^{-8}\ {\rm \Omega^{-1}m^{-1}}$ and
$1 \times 10^{-9}\ {\rm \Omega^{-1}m^{-1}}$. 
The variation in conductivity was due to two effects:
the conductivity is temperature dependent and the conductivity
decreased slowly in time. At a fixed temperature,
the main effect of the conductivity
drift is to shift the critical voltage, $V_c$, for the onset of
convection. The different temperatures were used to study the effect
of varying the material parameters and to offset the long term shifts
in $V_c$. By changing the temperature, $V_c$ was kept at
$\approx 11 {\rm V_{rms}}$. For all of the experiments reported
here, the drive frequency was 25 Hz.

Because of the shift in $V_c$, the following protocols were
followed for all of the experimental runs. All transitions are
reported in terms of $\epsilon = (V/V_c)^2 - 1$. To determine the
various transition points, the voltage was increased in steps of
$\Delta \epsilon$ ranging from $0.005$ to $0.01$ depending on the
precision of interest for a given run. For measurements of the
onset of the bimodal varicose instability ($\epsilon_{BV}$), a
single image was taken after the system was allowed to equilibrate
for 10 minutes. The relevant time scale is the director relaxation
time, which for this system is on the order of 0.2 s. As
$\epsilon$ is increased above $\epsilon_{BV}$, there is a Hopf
bifurcation to the oscillating bimodal varicose state at
$\epsilon \equiv \epsilon_{H}$. 
The value of $\epsilon_{H}$ was determined by quasi-statically
stepping $\epsilon$ from the bimodal varicose state. At each
step in $\epsilon$, a time series of images was taken.  The power
spectrum of the time series was computed, and the signature of the Hopf
bifurcation was the development of a nonzero frequency component.
For each run, $V_c$ was measured before and after the experiment
to determine the drift in $V_c$. The drift was a relatively constant
0.018 volts per hour, which corresponds to a shift in $\epsilon$
of 0.003 per hour. This level of drift in conductivity
did not adversely affect our ability to make comparison with theory
and was accounted for in all values of $\epsilon$ that are reported
here.

\section{Experimental Results}

Figure 2 shows four images of a typical pattern in the oblique
roll regime, and Fig.~3 shows four
images of a typical pattern above the bimodal varicose transition.
For each figure, the images are all of the same pattern and the following
protocols were used to take the images.
Image (a) has the polarizer and
analyzer parallel to each other and the undistorted director
(setup shown in Fig.~1a).
The $\lambda/4$ plate is orientated at $+45^\circ$
with respect to the undistorted director. In this case, one observes the
usual shadowgraph image. There is also an overall modulation of the
intensity due to the homogeneous twist mode.
For images (b), (c), and (d), the
polarizer has been rotated 90$^\circ$ with respect
to the undistorted director (setup shown in Fig.~1b).
Image (c) is a Fourier filtered version of image (b). A low-pass
Fourier filter was used to highlight the intensity variation
due to the homogeneous twist of the director. This long-wavelength
variation is difficult to detect in the raw image because
of the residual focusing effects
from the x-z distortion of the director.
For image (d), the optical axis of the $\lambda/4$ plate is orientated
-45$^\circ$ relative to the polarizer and the image is again
Fourier filtered with a low-pass filter. Therefore, image (d) should
be the complement of image (c). 

\begin{figure}[htb]
\epsfxsize = 3.2in
\centerline{\epsffile{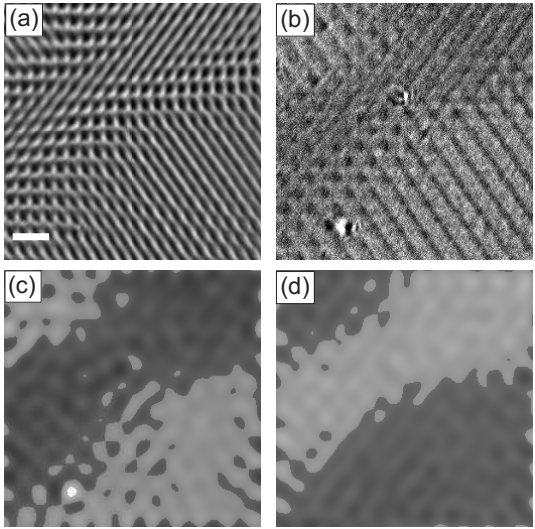}}
\caption{Four images of the oblique roll state in a
$0.7\ {\rm mm} \times 0.7\ {\rm mm}$ region of the cell. The
white bar in image (a) represents 0.1 mm.  Images (a) and
(b) are unprocessed. Images (c) and (d) have been processed
with a low-pass Fourier filter to highlight the effects of the
homogeneous twist of the director. The orientations of the
$\lambda/4$ plate and the polarizers are as follows.
(a) Unfiltered image with the polarizer aligned parallel
to the undistorted
director. (b) and (c) Unfiltered and filtered images,
respectively, with the polarizer aligned perpendicular
to the undistorted director. (d) Filtered image with the
polarizer
aligned perpendicular to the undistorted director and
the $\lambda/4$ plate rotated $90^\circ$ relative to
the orientation used to take image (b).}
\end{figure}

Both sets of images were obtained by jumping the voltage from
below $V_c$ to a value in the middle of the range for each
state. A jump was used to create a pattern with both the zig and
zag orientations to illustrate the different orientations of the
in-plane twist in a single image.
When the voltage is stepped slowly, a
single orientation of the rolls exists
over large regions of the cell.

With the orientation used for images (b) and (c),
the brighter regions correspond to regions of negative twist.
Therefore, Fig.~2b and c confirm that the twist orientation
is opposite the pattern wavevector,
as expected \cite{PP99}. For example, the region of zig rolls ({\bf q}
at $+\theta$) in the lower right corner appears brighter than
the region of zag rolls ({\bf q} at $-\theta$) diagonally across
the middle.
Also, Fig.~2d is clearly the complement of Fig.~2c, providing additional
evidence that the source of the variations is
an in-plane twist of the director. 

\begin{figure}[htb]
\epsfxsize = 3.2in
\centerline{\epsffile{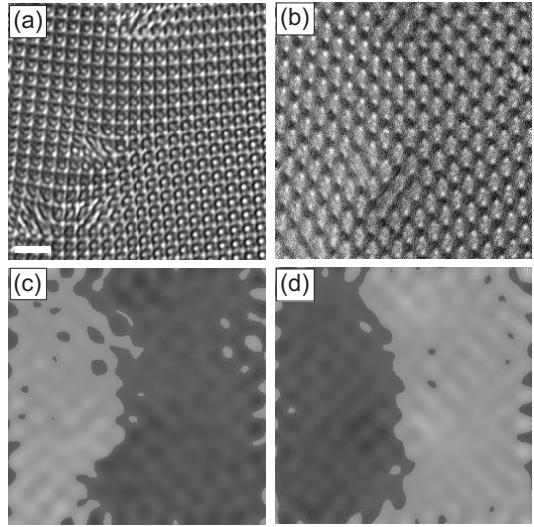}}
\caption{Four images of the bimodal varicose state in a
$0.7\ {\rm mm} \times 0.7\ {\rm mm}$ region of the cell. The
white bar in image (a) represents 0.1 mm.  Images (a) and
(b) are unprocessed. Images (c) and (d) have been processed
with a low-pass Fourier filter to highlight the effects of the
homogeneous twist of the director. The orientations of the
$\lambda/4$ plate and the polarizers are as follows.
(a) Unfiltered image with the polarizer aligned parallel
to the undistorted
director. (b) and (c) Unfiltered and filtered images,
respectively, with the polarizer aligned perpendicular
to the undistorted director. (d) Filtered image with the
polarizer
aligned perpendicular to the undistorted director and
the $\lambda/4$ plate rotated $90^\circ$ relative to
the orientation used to take image (b).}
\end{figure}

In Fig.~3, the bimodal varicose pattern is shown.
This state corresponds to the state SO2 in
Ref.~\cite{DCA98}. In general,
the bimodal varicose pattern is described by the superposition
of two modes with different amplitude and wavevectors, where one
of the wavevectors is of the zig-type and the other 
is of the zag-type.
This can be written as
$A \cos({\bf q}\cdot{\bf x}) + B \cos({\bf k}\cdot{\bf x})$, where
$A$ and $B$ are the two amplitudes of the modes.
In general, the two wavevectors are such that the modes are not
a degenerate zig and zag pair.
In this case, there are two degenerate
bimodal varicose states. One formed when the initial wavevector
{\bf q} is a zig state, and one
formed when {\bf q} is a zag state.  The two degenerate
states are shown in Fig.~3. The left
half of the image consists of a pattern where the
zig-type rolls were the initial state and zag-type rolls
grew as a result of the instability. The right half of the image is the
reverse case. In this case, the angle between $\bf q$ and
{\bf k} is 80$^\circ$.
It is this superposition of a zig-type and zag-type roll that
identifies the SO2 state as the bimodal varicose state.
Furthermore, Figs. 3c and 3d confirm that the direction
of the homogeneous twist
is still determined by the wavevector with
the maximum amplitude. This provides strong evidence for the
mechanisms described in Ref.~\cite{PP99} as the source of the
bimodal instability. Further work is needed to make quantitative
measurements of the change in the amplitude of the twist mode as
a function of the growth of the {\bf k} mode.

\begin{figure}[htb]
\epsfxsize = 3.2in
\centerline{\epsffile{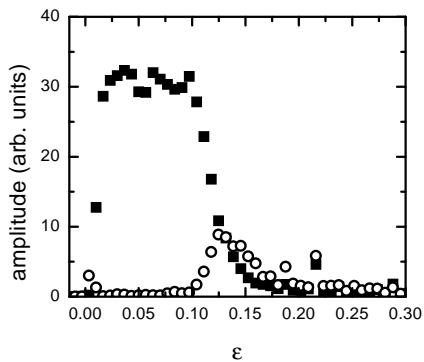}}
\caption{Amplitude of the pattern at the fundamental
wavevector {\bf q} (solid squares) and at
the dual wavevector {\bf k} (open circles) at an angle
of 85.3$^\circ$ with respect to {\bf q}. This shows the
bifurcation to oblique rolls at
$\epsilon = 0.007$ and the bimodal varicose instability
at $\epsilon = 0.1$. The initial small
amplitude at {\bf k} is due to the initial Hopf
bifurcation for which zig and zag coexist.
Within the resolution of these measurements,
the amplitude of the dual wavevector {\bf k}
grows continuously at the bimodal varicose instability.
See the text and Fig.~5 for a discussion of the
apparent decrease in the amplitude of both modes for large $\epsilon$.}
\end{figure}

Because the SO2 state corresponds to the bimodal varicose state,
measurements of the transition point ($\epsilon_{BV}$)
exist for some parameter values \cite{DCA98}. These values for
$\epsilon_{BV}$ are in very rough agreement
with calculations of the twist-mode
model \cite{P00}; however, detailed measurements of the transition
points have not been made. For example, in this system,
it has not been determined if the transition to the bimodal varicose
is forward or backward. In order to elucidate the nature of the
transition, Fig.~4 shows the amplitude of the pattern
at the wavevectors
{\bf q} and {\bf k} as a function of $\epsilon$ for the sample at
T = 45 $^\circ$C.  The amplitude of the pattern at a given wavevector
is calculated from the power
spectrum for each image. For each wavevector, the power is computed by
summing the power in a 3 x 3 square centered on the wavevector. The
amplitude is the square root of the power. Figure 4 provides strong
evidence that the bimodal
varicose instability is forward, within the resolution of
our measurements, because the amplitude of
the {\bf k} mode grows continuously from zero.

In Fig.~4, it appears that the amplitude of the {\bf q} mode
decreases at large $\epsilon$. Though leveling
off of the amplitude is expected, the decrease
is most likely an artifact
of  optical nonlinearities that arise at high $\epsilon$.
This is demonstrated in Fig.~5, which shows a typical image of
the bimodal varicose state 
at $\epsilon = 0.15$. Also shown in Fig.~5 is the
corresponding power spectrum. One clearly observes the large
number of peaks corresponding to nonlinearities in either
the optics or the pattern itself. In principle, one can compute
the amplitude of the director variation $A$ directly from the
images. In electroconvection, the shadowgraph images
contain contributions proportional to both $A$ and $A^2$. 
This adds some
complication for calculating amplitudes of superimposed oblique
rolls because the $A^2$ terms result in sums and differences of the
two wavevectors. However, the real problem for this
case is the additional
nonlinearities that result in the plethora of
diffraction peaks in Fig.~5b. One such additional problem is the
existence of caustics at these large amplitudes. 

\begin{figure}[htb]
\epsfxsize = 3.2in
\centerline{\epsffile{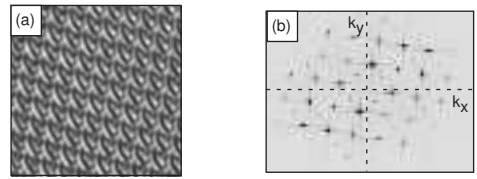}}
\caption{(a) Image of the pattern at $\epsilon = 0.15$. (b) Spatial
power spectrum of the image shown in (a).}
\end{figure}

An additional feature of the electroconvection in I52
is that the initial bifurcation is actually a continuous,
Hopf bifurcation to a state of superimposed zig and zag rolls.
The large jump in amplitude to the pure zig state at
$\epsilon = 0.01$ corresponds to the transition from
the initially traveling rolls to the stationary state. 
For all of the conductivities and temperatures
reported on here, there was an initial Hopf bifurcation. 

As $\epsilon$ is increased further, the bimodal varicose state
experiences a Hopf bifurcation to the oscillating bimodal varicose
state at $\epsilon \equiv \epsilon_{H}$.  
Figure 6 shows the power spectrum as a function of frequency
for the wavevectors {\bf q}, {\bf k}, and for the zero
wavevector of a typical time series above $\epsilon_{H}$.
The presence of a peak at finite $\omega$ is the signature
of the oscillating bimodal state. 
These power spectra were computed by taking a time series
of 64 images 0.5~s apart. The images covered a region containing
approximately 5 wavelengths of the pattern.
Each image was Fourier transformed,
and a time series of the Fourier transforms at each wavevector
of interest
was constructed. Then, the power spectra of each of these time series
was calculated. For these images, the arrangement of
polarizer, $\lambda/4$ plate, and analyzer described in
Fig.~1a was used. As discussed above,
the use of image subtraction implies that the power in the
zero wavevector corresponds to the amplitude of the twist mode, as
it represents a long wavelength variation of the intensity.
Also, for this state, the initial wavevector was of the
zag-type and the dual wavevector was of the zig-type. Therefore,
the twist rotation is positive, and with the optical setup
of Fig.~1a, this produces an overall increase of the image intensity.
Figure~6 illustrates two main points. First, the Hopf
bifurcation corresponds to an oscillation of the roll amplitudes
and the twist amplitude about their corresponding mean values (the
large peak at zero frequency).
Second, the oscillation amplitude is significantly less than
the mean value.

\begin{figure}[htb]
\epsfxsize = 3.2in
\centerline{\epsffile{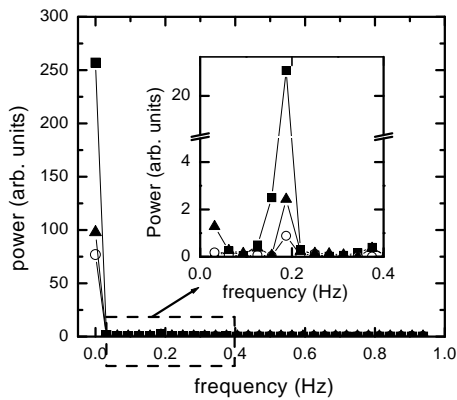}}
\caption{Power spectra from a time series of images at three
different wavevectors: zero wavevector (solid squares);
{\bf q} (solid triangles); and {\bf k} (open circles).
Here the initial wavevector {\bf q} was of the zag-type.
The full power spectrum shows the
large peak at zero frequency for each wavevector. This corresponds
to the mean of each mode. The insert shows a close up view of
the range indicated by the dashed box. The peak at approximately
0.2 Hz is the signature of the Hopf bifurcation from the
bimodal varicose state.}
\end{figure}

The oscillating bimodal varicose, as defined in Ref.~\cite{PR98b},
is a pattern of the form
$A(t) \cos({\bf q}\cdot{\bf x}) + B(t) \cos({\bf k}\cdot{\bf x})$, where
$A(t)$ and $B(t)$ oscillate roughly out-of-phase with each other and
around different mean values. For the oscillating state observed here,
Fig.~6 demonstrates that the mean values are different.
Figure 7 illustrates the behavior of the oscillating bimodal varicose
in real space and illustrates the out-of-phase nature of the oscillations.
The image in Fig.~7 is one of the individual frames used to
compute the power spectra shown in Fig.~6. Because of the optical
nonlinearities, the easiest way to directly observe out-of-phase oscillations
is to plot the local behavior of the pattern after Fourier filtering.
This is show in the plot in Fig.~7. The plot is constructed as follows.
For each image in the time series, the Fourier transform is computed.
Then, a Hanning window is applied to the region around the wavevector of
interest (and its complex conjugate). The inverse Fourier transform of
the result produces a real space image that corresponds to the mode
of interest. In this real space image, the pixels in a $2 \times 2$
region are averaged. The region is shown by the dot in the image in
Fig.~7. A time series of these average values is then constructed.
The plot in Fig.~7 is subset of the longer time series used to compute
the power spectra of Fig.~6.

Figure 7 clearly illustrates the oscillations of the twist-mode
(squares), zig-type mode (triangles), and zag-type mode (circles)
that comprise the oscillating bimodal varicose. The optical arrangement
of Fig.~1a was used while taking this time series. Therefore, the
positive amplitude of the twist-mode represents a positive 
rotation angle.
This is consistent with the dominant mode being of the
zag-type. Also, an
increase in the zero wavevector intensity represents an increase
in the twist-mode amplitude, which is correlated with the increase
in the zag mode amplitude.  
A mechanism for the oscillating bimodal varicose is
discussed in Ref.~\cite{PP99}, and the evidence provided by Figs.~6 and
7 for this mechanism is discussed in Sec. IV.

\begin{figure}[htb]
\epsfxsize = 3.2in
\centerline{\epsffile{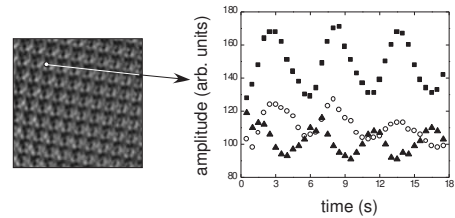}}
\caption{Time series of the amplitude of the three modes
whose power spectra are plotted in Fig.~6. The time series
is the average amplitude of a region 2 pixels by 2 pixels that
is shown in the image by the white dot. The three modes are
identified by their wavevector: zero wavevector (squares),
{\bf q} (circles), and {\bf k} (triangles).}
\end{figure}

\section{Discussion}

The results presented in this paper highlight three main points.
First, the twist mode is present in the oblique roll states in I52,
as it should be. Because it is a weakly damped mode \cite{PP99},
its amplitude is actually relatively large. Therefore,
it must be included in any extended weakly
nonlinear description
of the system.

Second, the observed qualitative features of
the oscillating bimodal state are consistent with the mechanism
proposed in Ref.~\cite{PP99}. In particular, the results presented
in Fig.~7 provide strong evidence that the twist
mode is responsible for the dynamics of this state.
The plot in Fig.~7 should be compared to the analogous plot in
Fig.~8 of Ref.~\cite{PP99}. In both cases, the amplitudes 
of the zig and zag-type rolls are out-of-phase with each other.
The twist-mode is found to oscillate with a phase intermediate
to the two modes. This is strong evidence for the
proposed director-wavevector frustration mechanism where the
the oscillations of the
twist mode mediate the oscillations in the
{\bf q} and {\bf k} modes \cite{PP99}.

Finally, the WEM effects may remain important in the nonlinear
states and have an effect on the location of
the bimodal varicose transition and the
following Hopf bifurcation. The fact that the WEM effects are present
is clearly shown by the initial Hopf bifurcation that is present at all
the parameters used here. The calculations in Ref.~\cite{PP99}
that include the twist amplitudes are based on the standard model of
electroconvection, which do not allow for the observed primary
Hopf bifurcation. However,
the fact that the traveling waves were
always replaced with a stationary pattern at $\epsilon \leq 0.01$
suggests that the WEM effects are
``weak'' in some sense. Therefore, it is not unreasonable to 
qualitatively compare
the results presented here with calculations based on the twist-mode
model. However, as is shown in Table I, there is some
quantitative disagreement between the twist-mode model
and the experiment.

\begin{table} {TABLE I. Comparison of theoretical and
experimental values for two different temperatures.}
\begin{tabular}{ccccc}
Temperature ($^\circ$C) & $\theta$ & $\alpha$ & $\epsilon_{BV}$
& $\epsilon_{H}$ \\ \hline
42 (exp.) & 32.0$^\circ$ & 82.4$^\circ$ & 0.079 & 0.17 \\ 
42 (theory) & 32.0$^\circ$ & 82$^\circ$ & $0.08^*$ & none \\
55 (exp.) & 37.9$^\circ$ & 86.3$^\circ$ & 0.03 & 0.19 \\
55 (theory) & 37$^\circ$ & 80$^\circ$ & 0.06 & none \\  
\end{tabular}
{$^*$A maximum in the growth rate of the dual is found, but the growth
rate is still negative.}
\end{table}

Table I shows a comparison between calculations \cite{P00}
and experiment for
two temperatures for the following measured quantities:
the initial angle of the rolls with respect to the undeformed
director ($\theta$); the angle $\alpha$ between 
{\bf q} and {\bf k} at the bimodal varicose instability;
the transition point to the bimodal varicose state
($\epsilon_{BV}$); and the subsequent Hopf bifurcation to
the oscillating bimodal varicose ($\epsilon_{H}$). There is good
agreement between $\theta$ and $\alpha$. The agreement for
$\theta$ is not surprising because the
WEM predicts only a small shift in $\theta$ from the standard model
value. Likewise, the agreement for $\alpha$ is not surprising
because the WEM does not appear to shift angles significantly.
The asterisk next to the calculated value of
$\epsilon_{BV}$ for the T = 42 $^\circ$C case indicates that
there is a maximum in the growth rate of {\bf k} at this point,
but the growth rate is still negative. In fact, for this
temperature, the growth rate of {\bf k} does not become
positive within the twist-mode model. Also, for the
T = 55 $^\circ$C case, the calculation predicts a value
for $\epsilon_{BV}$ that is too large.
For all the parameter values used in the experiment, a transition
to the oscillating bimodal varicose from the bimodal varicose state
was observed. Over this same range of parameters, the calculations
predict a restabilization of a single roll state that supersedes
the oscillating bimodal varicose.

In addition to possible WEM effects,
there are a number of additional sources for the above outlined
quantitative disagreements between experiment and the twist-mode
model. First, the
locations of $\epsilon_{BV}$ and $\epsilon_{H}$
have not yet been calculated
in a fully nonlinear calculation and can only be estimated within
the context of the extended weakly nonlinear analysis. Second,
there are issues of pattern selection that are not addressed in
the weakly nonlinear analysis. For example, even though the
weakly nonlinear analysis predicts the restabilization of the
oblique rolls, the pattern may still select the oscillating
bimodal varicose state. Finally,
there remains important uncertainties in the material parameters
of I52 that make quantitative comparison between
theory and experiment difficult.

The confirmation of the existence of the twist mode in
electroconvection in I52 has important consequences for both
the states of spatiotemporal chaos \cite{DAC96a}
and the localized worm states \cite{DAC96b}
that have been observed in this
system. Because both of these states involve the superposition of
oblique rolls, the twist mode must automatically be present,
as observed here.
In principle, if the states were a superposition of equal amplitudes
of zig and zag, the twist mode would have zero amplitude as the
two set of rolls produce opposite twists. However, for the state
of spatiotemporal chaos, the amplitudes vary irregularly.
Therefore, the twist mode may play an active roll in the dynamics.
Likewise, in the worm state, the zig and zag rolls have different
amplitudes at the edges of the worm. Therefore, the twist mode may
play a role in the localization mechanism of the worms as an
additional slow field. 

\acknowledgments

I thank Emmanuel Plaut and Werner Pesch for useful discussions
Also, I thank Emmanuel Plaut for providing me
with unpublished theoretical calculations. This work was
supported by NSF grant DMR-9975479.

\end{document}